\documentclass[aps,prl,twocolumn,showpacs,preprintnumbers,superscriptaddress]{revtex4}
\pdfoutput=1
\usepackage{bm,amsmath,amssymb,latexsym,mathrsfs,enumerate,color}
\usepackage[mathcal]{euscript}
\usepackage[hypertex]{hyperref}
\usepackage{graphicx}
\usepackage{sidecap}

\renewcommand{\vec}[1]{\bm{#1}}
\begin{document}

\title{Can immobile magnetic vortex nucleate a vortex--antivortex pair?}

\author{Volodymyr P. Kravchuk}
 \email[Corresponding author. Electronic address:]{vkravchuk@bitp.kiev.ua}
 \affiliation{Bogolyubov Institute for Theoretical Physics, 03143 Kiev, Ukraine}

\author{Yuri Gaididei}
 \affiliation{Bogolyubov Institute for Theoretical Physics, 03143 Kiev, Ukraine}

\author{Denis D. Sheka}
     \affiliation{National Taras Shevchenko University of Kiev, 03127 Kiev, Ukraine}

\date{\today}

%
%

\begin{abstract}
It is shown that under the action of rotating magnetic field an
\emph{immobile} vortex, contrary  to a general belief, can nucleate
a vortex-antivortex pair and switch its polarity. Two different
kinds of \textsf{OOMMF} micromagnetic modeling   are used: (i)  the
original vortex is pinned by the highly--anisotropic easy-axes
impurity at the disk center, (ii)  the vortex  is pinned by
artificially fixing the magnetization inside the vortex core in  the
planar vortex distribution. In both types of simulations a dip
creation  with a consequent vortex-antivortex pair nucleation is
observed. Polarity switching occurs  in the former case only. Our
analytical approach is based on the transformation to the rotating
frame of reference, both in the real space and in the magnetization
space, and on the observation that the physical reason for the dip
creation is softening of the dipole magnon mode due to magnetic
field rotation.

\end{abstract}

\pacs{75.10.Hk, 75.40.Mg, 05.45.-a, 72.25.Ba, 85.75.-d}



\maketitle

\begin{figure*}
\includegraphics[width=\textwidth]{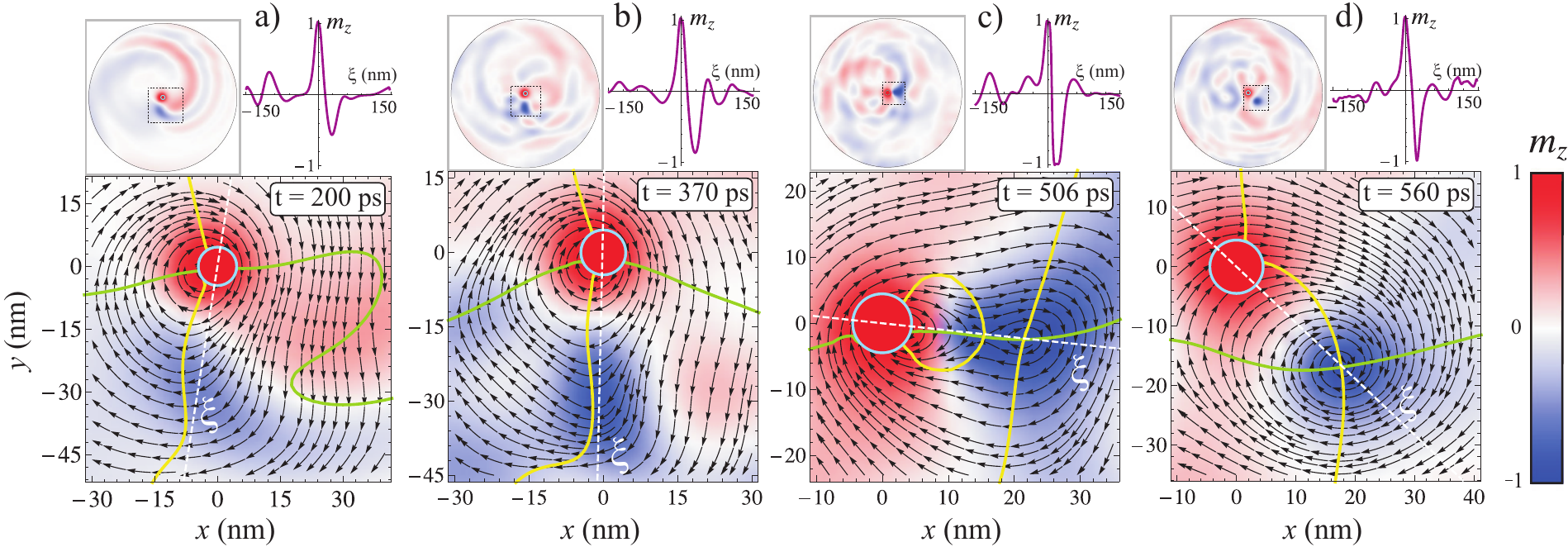}
\caption{(Color online). The modeling of the vortex--antivortex pair
formation for the pinned vortex on the impurity in the Py disk
($300$~nm diameter, $20$~nm thickness, $9$~nm is a diameter of impurity with
the easy--axis anisotropy $K=17$~MJ/m$^3$).
Snapshots describe the magnetization dynamics at different times under
 the influence of the rotating in--plane magnetic field ($\omega=6$~GHz, $B=20$~mT).
 The lower row shows the in--plane magnetization dynamics near the disk center.
 The in-plane magnetization distribution is indicated by streamlines with arrows.
 Green and yellow solid lines corresponds to the isosurfaces $m_x=0$ and $m_y=0$, respectively.
 The out--of--plane magnetization profiles along the line $\xi$ crossing the impurity is shown above the snapshots
 (right pictures). Left pictures indicate the in--plane magnetization dynamics in the whole disk.
}%
\label{fig:imp}
\end{figure*}

A magnetic vortex forms a ground state of submicron sized magnetic
particles, which provides high density storage and high speed
magnetic RAM \cite{Cowburn02}. One bit of information corresponds to
the upward or downward magnetization of the vortex core (vortex
polarity). Exciting the vortex motion by a high-frequency magnetic
fields or by a spin polarized currents, one can switch the vortex
polarity on a picoseconds time scale. It is known that the switching
process is mediated by a vortex-antivortex pair creation
\cite{Waeyenberge06}. Typically to observe this switching
phenomenon, one has to excite the low frequency gyroscopical mode,
which causes the vortex gyromotion. To the best of our knowledge the
vortex core switching was always observed experimentally and by
simulations only for a moving vortex. Moreover, there is a strong
belief that the switching occurs `whenever the velocity of
vortex-core motion reaches its critical velocity' $v_{\text{cri}}$
\cite{Lee08c}, which is determined only by the exchange constant
$A$; for typical soft materials is about $v_{\text{cri}}\sim300$~m/s
\cite{Lee08c,Vansteenkiste08a}.

In the present Letter we predict the switching for the
\emph{immobile} vortex by a rotating magnetic field. The switching
picture also involves the mechanism of a dip formation followed by a
nucleation of vortex--antivortex pair. We found that the driving
force for the dip formation is  the in--plane magnetization
inhomogeneity, created by a vortex together with the rotating
magnetic field; the vortex out--of--plane structure and the vortex
velocity are not principle for this mechanism. These conclusions are
confirmed by the micromagnetic simulations and by a simple
analytical picture.

Recently, we have reported about the vortex core switching by the
homogeneous rotating field $\vec{B}(t) = \left(B\cos\omega t,
B\sin\omega t, 0\right)$ in the ten GHz range \cite{Kravchuk07c},
which is much higher than the gyrofrequency and is in  the frequency
range of the higher azimuthal mode with $|m|=1$
\cite{Ivanov05,Zaspel05,Zivieri08}. Opposite to the pumping with the
low gyrofrequency, which results in the visible gyroscopic motion of
the vortex position, the high frequency field leads to small
amplitude (few nm) oscillations of the vortex position, see the
trajectories on Fig.~3 of Ref.~\cite{Kravchuk07c}. Typical
velocities when switching occurs are lower than reported in
Ref.~\cite{Lee08c}. There appears a question: is it necessary for
vortex to move in order to switch its polarity?

To achieve switching of the immobile vortex we have performed
numerically two different kinds of modeling using  \textsf{OOMMF}
micromagnetic simulations \footnote{We used for all \textsf{OOMMF}
simulations material parameters adopted for the Py particle: the
exchange constant $A=2.6\times10^{-6}$ erg/cm, the saturation
magnetization $M_S=8.6\times10^{2}$ G, the damping coefficient $\eta
= 0.006$ and the anisotropy was neglected. This corresponds to the
exchange length $\ell = \sqrt{A/4\pi M_S^2} \approx 5.3$nm. The mesh
cells have sizes $3\times3\times h$ nm, where $h$ is thickness of
the sample. The applied field was used in form $\vec
b(t)=b(1-e^{-t^2/\Delta t^2})(\cos\omega t,\sin\omega t,0)$, with
$\Delta t=50$ps.}

(i) We have pinned the vortex by the highly--anisotropic easy-axes
impurity at the disk center, see Fig.~\ref{fig:imp}. Since the
magnetization is always hold perpendicular to the disk plane within
the impurity, shifting of the vortex core from the impurity leads to
the increasing of magnetostatic energy of the system. Therefore one
can use such type of impurity for pinning of a vortex with
out-of-plane core structure in conditions of a weak external
influence. Initially, the vortex has a positive polarity $p=+1$. By
applying the ac field, which rotates against the vortex polarity,
$\omega p<0$ [in a clockwise (CW) direction in our case], one can
easily resolve the formation of the dip near the pinned vortex, see
Fig.~\ref{fig:imp}(a). Under the action of the field the dip
deepens, see  Fig.~\ref{fig:imp}(b). When the amplitude of the dip
reaches its maximum value, a vortex--antivortex pair is created, see
Fig.~\ref{fig:imp}(c). The positions of the vortices and the
antivortex can be identified by the cross--section of isosurfaces
$m_x=0$ and $m_y=0$ \cite{Hertel06}. One can identify from
Fig.~\ref{fig:imp}(c) positions of new born vortex and antivortex as
well as the position of the initial vortex, which is (as it is seen
from the figure) still at origin. The further dynamics is well known
\cite{Waeyenberge06,Hertel06,Hertel07,Gaididei08b}: the new born
antivortex annihilates with pinned original vortex, and eventually
only  the new born vortex with the opposite polarity survives (see
Fig.~\ref{fig:imp}(d)).

\begin{figure*}
\includegraphics[width=\textwidth]{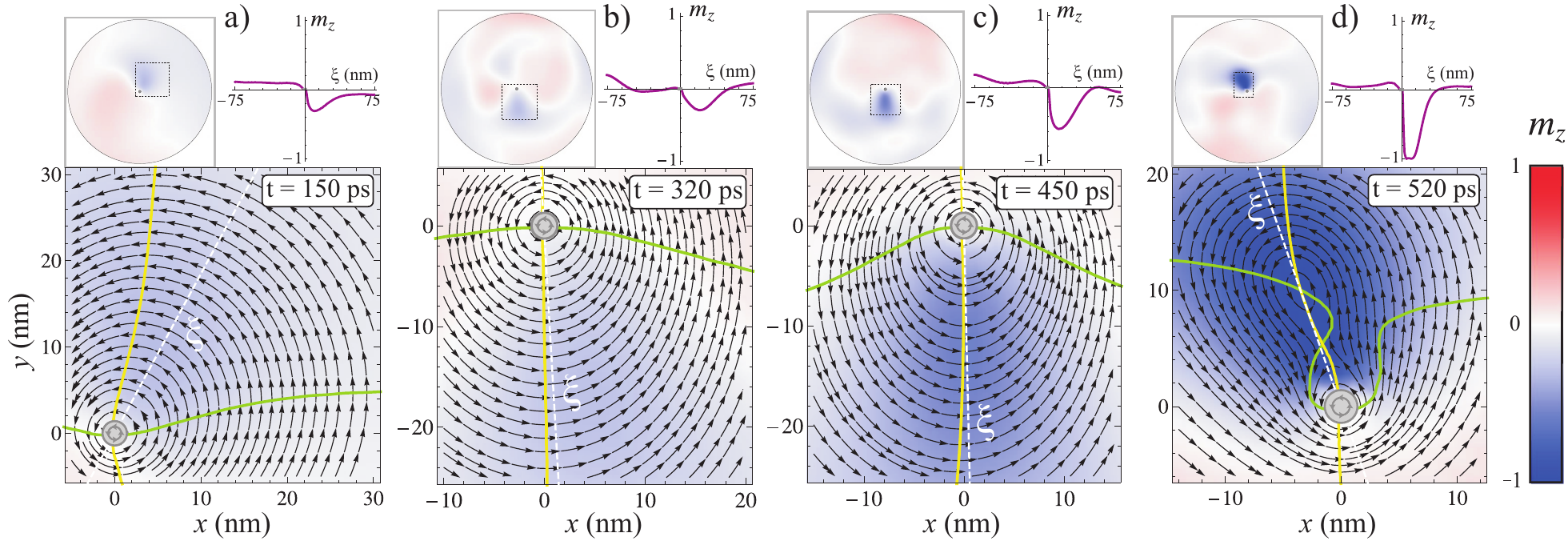}
\caption{(Color online). The vortex--anivortex pair formation in a
Py disk ($150$~nm diameter, $20$~nm thickness) within the fixed core
modeling (see the details in the text) under the CW field influence ($\omega=8$~GHz, $B=20$~mT).
Notations are the same as on Fig.~\ref{fig:imp}.}%
\label{fig:fixed}
\end{figure*}

(ii) In the second kind of simulations we artificially pinned the
vortex core by fixing the magnetization inside the vortex core in
the planar vortex distribution, where the normalized magnetization
$\vec{m}=\left(\sqrt{1-m_z^2}\cos\phi;
\sqrt{1-m_z^2}\sin\phi;m_z\right)$ takes a form $m_z=0$ and
$\phi=\chi\pm\pi/2$ with $(r,\chi)$ being the polar coordinates in
the disk plane. By this artificially pinned vortex state
configuration, the central vortex is \emph{a priori} immobile, and
moreover, without out--of--plane structure.  The direction of the
dip magnetization  is determined only by the direction of the field
rotation: if we apply a CCW (CW) field, there appears a dip with
$m_z>0\,\,(m_z<0)$, see Fig.~\ref{fig:fixed}. This high frequency
field excites also magnon modes. One can resolve the mode with $m=1$
on Figs.~\ref{fig:fixed}(a) and (c), and the mode with $m=2$ on
Fig.~\ref{fig:fixed}(b). During the pumping, the amplitude of the
dip increases, and finally, the vortex--antivortex pair is created,
see Fig.~\ref{fig:fixed}(d). Note that the new--born antivortex can
not annihilate with the initial vortex in the fixed core model: it
either annihilates with the new born vortex and the scenario repeats
again and again, or it stands near the original fixed vortex, while
the new born vortex is involved by the field.

\begin{figure}
\includegraphics[width=0.75\columnwidth]{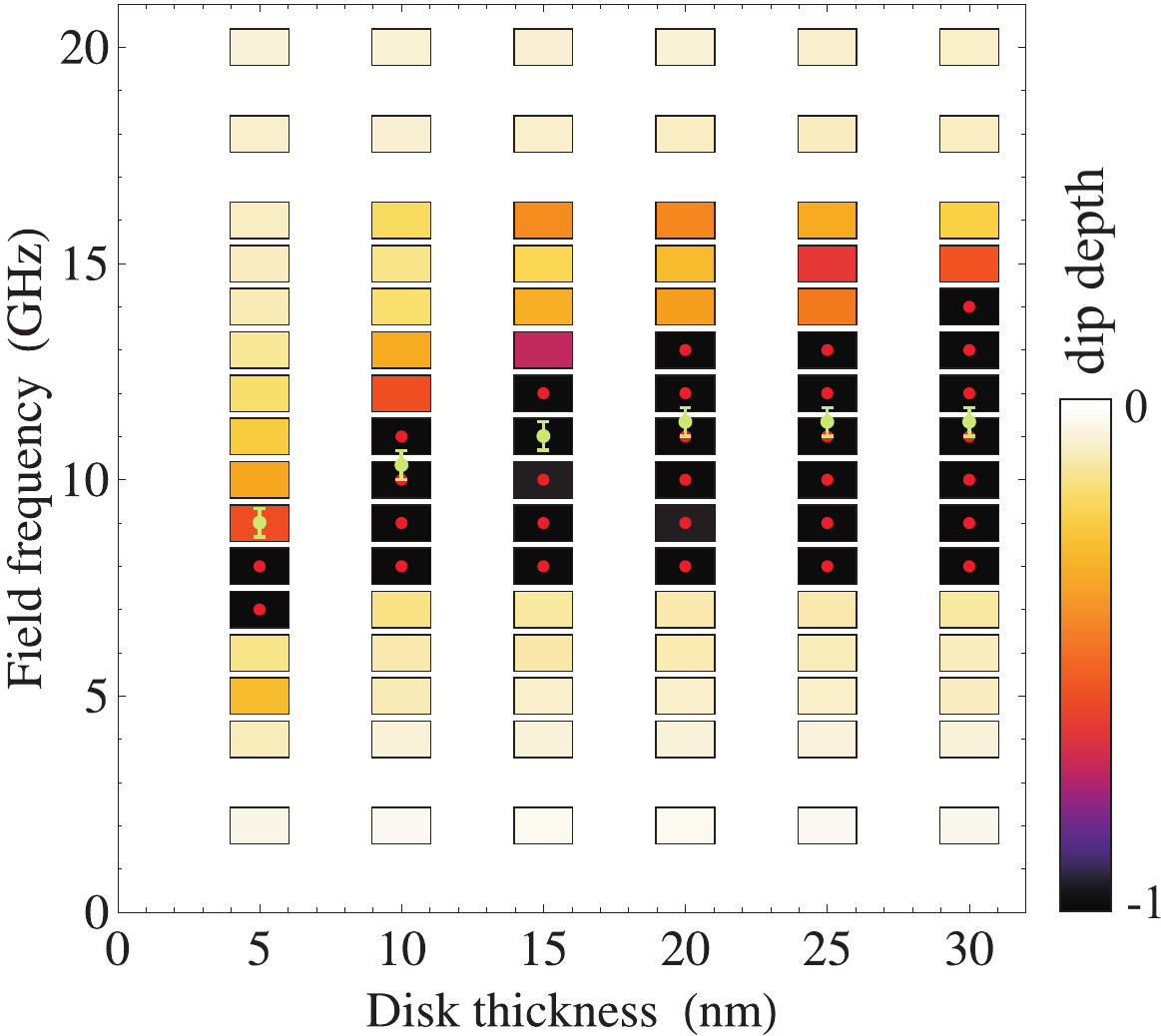}
\caption{(Color online) Diagram of the frequency range,
where the dip formation occurs for a fixed diameter disks
($2R=150$~nm) under the influence of a fixed field amplitude
($B=20$~mT) rotating field. Colors indicate the depth of the dip.
The circle over black symbol indicates that the vortex-antivortex
pair creation occurs. Light--green symbols indicate frequencies of
the azimuthal eigenmode with $m=1$.}%
\label{fig:diagram}%
\end{figure}

\begin{figure}
\includegraphics[width=0.95\columnwidth]{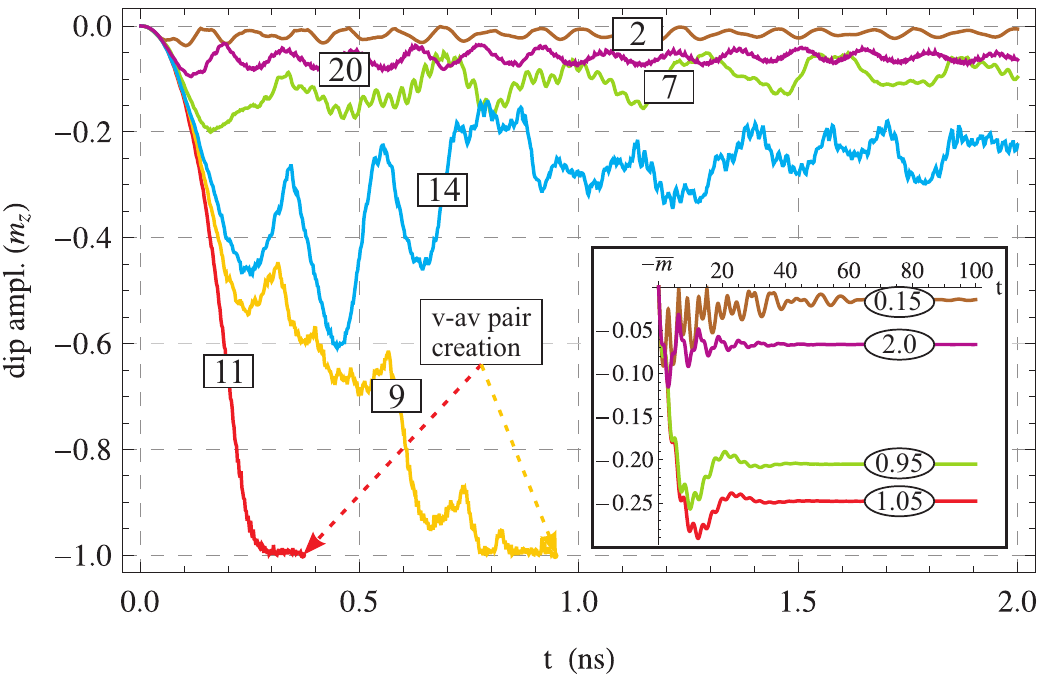}
\caption{(Color online) Amplitude of $m(t)$ at the minimum (dip center) for the disk($2R=150$~nm, $h=15$~nm) under the influence of rotating field with a fixed amplitude $B=20$~mT and different frequencies.
Numbers in rectangle frames denotes frequencies in GHz.
The inset contains the time dependence of the dip depth $-\bar{m}(t)$ obtained from the effective Lagrangian (\ref{eq:L-eff}) for $\omega_0=2,\omega_1=1, b_e=0.1,\eta=0.1, c_1=c_2=0.25,c_3=0.1$. Corresponding frequencies in dimensionless units are shown in oval frames.}%
\label{fig:dip_vs_t}%
\end{figure}

The goal of our study is to reveal the mechanism of the dip
formation. We follow the simplest model of fixed core planar vortex.
Numerically, we found that the dip formation occurs in well defined
range of the field parameters $(\omega,B)$, see
Fig.~\ref{fig:diagram}. When the frequency of the field is less than
a critical value $\omega_{c1}$, the dip is not formed. The same
happens for the high frequency field ($\omega>\omega_{c2}$). More
accurate, out of the borders of the range
($\omega_{c1},\,\omega_{c2}$), the depth of the dip rapidly
decreases, what is indicated by colors on Fig.~\ref{fig:diagram}.
Both critical values $\omega_{c1}$ and $\omega_{c2}$ depend on the
disk thickness, which counts in favor of nonlocal
magnetostatic nature of this phenomenon. An example of the temporal
dependence of the dip depth is plotted on Fig.~\ref{fig:dip_vs_t} for a certain disk thickness (15nm), for which
$\omega_{c1}$=8GHz and $\omega_{c2}$=12GHz. For $\omega<\omega_{c1}$
just as for $\omega>\omega_{c2}$ the dip depth does not reach its
minimal value $m_z=-1$, unlike the case
$\omega_{c1}<\omega<\omega_{c2}$, when the vortex-antivortex pair is
born. The depth of the dip usually makes a number of oscillations
while it reaches the limit value. Only for frequencies, which are
close to the eigenfrequency of the mode with $|m|=1$, it changes
monotonously.

The magnetic energy of the system under consideration consists of
three terms:
$\mathscr{E}=\int\!\mathrm{d}^3\vec{r}\left(W_{\text{ex}}+W_{\text{ms}}+W_{\text{f}}\right)$,
where $W_{\text{ex}}=\ell^2\left(\nabla\vec{m}\right)^2$ is the
exchange energy
density,$W_{\text{ms}}=\frac{1}{8\pi}\int\mathrm{d}^3\vec
r'\bigl(\vec m(\vec r)\cdot\nabla\bigr)\bigl(\vec m(\vec
r')\cdot\nabla'\bigr)|\vec r-\vec r'|^{-1}$
 is the
density of magnetostatic interaction energy. And
$W_{\text{f}}=-b\sqrt{1-m_z^2}\cos(\phi-\omega t)$ is the
interaction with a magnetic field $\vec{b}=\left(b\cos\omega
t,b\sin\omega t,0\right)$. Here and below normalized quantities are
used: $\mathscr{E}=E/(4\pi M_s^2)$, $b=B/(4\pi M_s)$, $\omega$ and
$t$ are measured in units $\Omega$ and $1/\Omega$ respectively,
where $\Omega=4\pi\gamma M_s$ and $\gamma$ is gyromagnetic ratio. In
the no-driving case $b=0$ the magnetic energy $\mathscr{E}$ of
\emph{cylindrical} nanoparticles  is invariant under transition into
a rotated frame of reference: $\tilde\chi=\chi-\alpha$,
$\tilde\phi(\tilde\chi)=\phi(\chi)-\alpha$. This means that the
quantity $J=M_z+L_z$ with $M_z=\int\!\mathrm{d}^3\vec{r}(1-m_z)$
being the magnetization along the cylindrical axis $z$ and
$L_z=-\int\!\mathrm{d}^3\vec{r}(1-m_z)\partial_{\chi}\phi$ being the
$z$-component of the orbital momentum, is conserved.  When the ac
field is turned on, the total momentum $J$ is not conserved anymore
but its existence allows to obtain that in the rotating frame of
reference $\tilde{\chi}=\chi-\omega t$, $\tilde r=r$, $\tilde t=t$,
$\tilde\phi(\tilde\chi)=\phi(\chi=\tilde\chi+\omega t)-\omega\tilde
t$ the magnetic energy  is time-independent  and has the form
\begin{equation*} \label{eq:magn_energy}
\tilde{\mathscr{E}}=\mathscr{E}-\omega J,
\end{equation*}
where the last term represents the rotation energy. The dynamics of
the system is governed by the Landau--Lifshitz--Gilbert equations
which in the rotating frame of reference have the form
\begin{equation*} \label{eq:LLG}
\frac{\delta \mathscr{L}}{\delta
m_z}+\frac{\eta}{1-m_z^2}\frac{\delta \mathscr{E}}{\delta
\tilde{\phi}}=0,\qquad\frac{\delta \mathscr{L}}{\delta
\tilde{\phi}}+\eta\,(1-m_z^2)\,\frac{\delta \mathscr{E}}{\delta m_z}
= 0,
\end{equation*}
where
$\mathscr{L}=(1+\eta^2)\int\!\mathrm{d}^3x(m_z-1)\partial_t\tilde{\phi}-\tilde{\mathscr{E}}$
is the Lagrangian and $\eta$ is  a damping constant.


To gain some insight how  the interaction with the magnetic field
(which is static in the rotating frame) together with the rotation
provides the dip creation we use the Ansatz
\begin{equation} \label{eq:Ansatz}
\begin{split}
m_z&=\alpha_0(t)f_0(r)+ \left[\alpha_1(t)\cos\chi+\alpha_{-1}(t)\sin\chi\right]f_1(r),\\
\phi&=\chi+\frac{\pi}{2} +\beta_0(t)g_0(r)\\
&+ \left[\beta_1(t)\cos\chi+\beta_{-1}(t)\sin\chi\right]g_1(r),
\end{split}
\end{equation}
where $\bigl(f_m(r), g_m(r)\bigr)e^{i m \chi}$ are the magnon
eigenfunctions  with the eigenfrequencies  $\omega_m\,(m=0,1)$ on
the planar vortex background  (see {\it e.g.} \cite{Ivanov02a} for
general case) and $\alpha_m,$ and  $\beta_m,$ are time dependent
coefficients. Inserting the Ansatz (\ref{eq:Ansatz}) into the
Lagrangian $\mathscr{L}$ and carrying out integrations over the
spatial coordinates, we get  an effective Lagrangian in the form
\begin{equation} \label{eq:L-eff}
\begin{split}
\mathscr{L}^{\text{eff}}&=\sum_{m=0,\pm 1}\Big(\alpha_m\dot{\beta}_m -m\,\omega\, \alpha_m\,\beta_{-m}\Big)-\mathscr{E}^{\text{eff}},\\
\mathscr{E}^{\text{eff}}&=\frac{1}{2}\sum_{m=0,\pm 1}\,\omega_{|m|}\,\left(\alpha_m^2+
 \beta_m^2\right) - b_e\beta_1\\
&+c_0\,\alpha_0^4+c_1\alpha_0 \left(\alpha_1\beta_{-1}-
\alpha_{-1}\beta_1\right)+c_2 \left(\alpha_1^2 +\alpha_{-1}^2\right)^2\\
&+
c_3\left(\alpha_1^2+\alpha_{-1}^2\right)\left(\beta_1^2+
\beta_{-1}^2\right)+c_4\Big[\Big(\alpha_1\beta_1+\alpha_{-1}\beta_{-1}\Big)^2
\\
&-\Big(\alpha_{-1}\beta_1-\alpha_1\beta_{-1}\Big)^2
\Big]+c_5\alpha_0^2\left(\alpha_1^2+\alpha_{-1}^2\right)\\
&+
c_6\alpha_0^2\left(\beta_1^2+\beta_{-1}^2\right)+c_7\alpha_0\beta_0
\left(\alpha_1\beta_{1}-
\alpha_{-1}\beta_{-1}\right)\\
&+c_8\beta_0^2
\left(\alpha_1^2+\alpha_{-1}^2\right)+\Big(\text{higher order terms}\Big).
\end{split}
\end{equation}
Here the coefficients $c_j$ with $j=\overline{0,\dots,8}$ are due to
nonlinear terms in the magnetic energy $\mathscr{E}$ and $b_e$ is an
effective strength of the magnetic field. Equations of motion for the parameters $\alpha_m,~\beta_m$ have the form
\begin{eqnarray}\label{eqab}\dot{\alpha_m}=m\,\omega\,\alpha_{-m}-
\frac{\partial \mathscr{E}^{\text{eff}}}{\partial\beta_m}-
\eta\frac{\partial \mathscr{E}^{\text{eff}}}{\partial\alpha_m},\nonumber\\
 \dot{\beta_m}=m\,\omega\,\beta_{-m}+
\frac{\partial \mathscr{E}^{\text{eff}}}{\partial\alpha_m}-
\eta\frac{\partial
\mathscr{E}^{\text{eff}}}{\partial\beta_m}.\end{eqnarray}Thus the
problem under consideration is reduced to the set of three
nonlinearly coupled oscillators. When $\omega\ll\omega_1$ the linear
terms in  Eq. (\ref{eq:L-eff}) dominate and the out-of-plane
component of the magnetization
$\bar{m}=\sqrt{\alpha_0^2+\alpha_1^2+\alpha_{-1}^2}$ is small, see
inset in Fig.~\ref{fig:dip_vs_t}. However,   as a result of rotation
the magnon frequencies in the rotating frame of reference are
shifted, $\tilde{\omega}_m = \omega_m-m\omega$, and the  dipole
magnon mode (\emph{i.e.} the azimuthal mode with $m=1$) becomes
soft, which can be identified from Fig.~\ref{fig:diagram}. Near the
threshold where $\tilde{\omega}_1$ is small, the nonlinear terms in
the Lagrangian $\mathscr{L}$ become crucial and the system goes to a
new regime with the finite value of $\bar{m}$, see the inset in
Fig.~\ref{fig:dip_vs_t}. The appearance of the finite $\bar{m}$ we
identify with the dip creation. The dip direction is determined by
the product $b_e\,\omega$: when $b_e\,\omega>0(<0)$ the
out--of--plane component of magnetization is negative (positive).
This result is in a full agreement with the results of full--scale
numerical simulations.

In conclusion, we showed that in the presence of {\it immobile}
planar vortex the rotating magnetic field produces a
vortex-antivortex pair. This process is the most efficient in a
finite frequency interval which is determined by the aspect ratio
and material properties of the nanoparticle. The physical reason of
the dip creation with a consequent vortex-antivortex nucleation is
softening the dipole magnon mode and magnetic field induced breaking
cylindrical symmetry in the rotating frame of reference. In some
respects such softness of modes due to rotation is similar to the
problem of instability of BEC under rotation \cite{Isoshima99}, and
to the Zel$'$dovich--Starobinsky effect for a rotating black hole
\cite{Takeuchi08}. Under the continuous pumping, the soft mode is
excited, and the system goes to the nonlinear regime, which results
in the dip formation.

The authors thank H.~Stoll and B.~Van~Waeyenberge for helpful
discussions. The authors thank the MPI Stuttgart, where part of this
work was performed, for kind hospitality and acknowledge the support
from Deutsches Zentrum f{\"u}r Luft- und Raumfart e.V.,
Internationales B{\"u}ro des BMBF in the frame of a bilateral
scientific cooperation between Ukraine and Germany, project
No.~UKR~08/001.


\end{document}